\documentclass[twocolumn,showpacs,superscriptaddress,nofootinbib,prd,floatfix]{revtex4}

\newcommand{\Lo}{{\mathcal L}}

\usepackage{graphicx}
\usepackage{color}
\usepackage{amsmath}
\usepackage{subfigure}
\usepackage{verbatim}

\begin{document}

\title{The harmonic structure of generic Kerr orbits}

\author{Rebecca Grossman}
\email{becky@phys.columbia.edu}
\affiliation{Physics Department, Columbia University, New York, NY
  10027}
\author{Janna Levin}
\email{janna@astro.columbia.edu}
\affiliation{Department of Physics and Astronomy, Barnard College of
  Columbia University, 3009 Broadway, New York, NY 10027}
\affiliation{Institute for Strings, Cosmology and Astroparticle
  Physics (ISCAP), Columbia University, New York, NY 10027}
\author{Gabe Perez-Giz}
\email{gabe@phys.columbia.edu}
\affiliation{Physics Department, Columbia University, New York, NY
  10027}

\begin{abstract}
\label{sec:abs}

Generic Kerr orbits exhibit intricate three-dimensional motion. We
offer a classification scheme for these intricate orbits in terms of
periodic orbits. The crucial insight is that for a given effective
angular momentum $L$ and angle of inclination $\iota$, there exists a
discrete set of orbits that are geometrically $n$-leaf clovers in a
precessing {\it orbital plane}. When viewed in the full three
dimensions, these orbits are periodic in $r-\theta$. Each $n$-leaf
clover is associated with a rational number,
$1+q_{r\theta}=\omega_\theta/\omega_r$, that measures the degree of
perihelion precession in the precessing orbital plane. The rational
number $q_{r\theta}$ varies monotonically with the orbital energy and
with the orbital eccentricity.  Since any bound orbit can be
approximated as near one of these periodic $n$-leaf clovers, this
special set offers a skeleton that illuminates the structure of all
bound Kerr orbits, in or out of the equatorial plane.
\end{abstract}


\maketitle

\section{Introduction}
\label{sec:intro}

Black hole orbits are defined by precession. The perfectly closed
ellipse of Kepler's Laws gives way to the relativistic precession of
Mercury's perihelion in the weak field around a star. In the
strong-field, perihelion precession in the equatorial plane of a black
hole can result in zoom-whirl orbits for which the precession is so
great at closest approach that the particle executes multiple circles
before falling out to apastron again. An orbit out of the equatorial
plane, the plane perpendicular to the spin axis of the black hole, is
shaped by yet another kind of precession -- precession of the orbital
plane. These most general black hole orbits live in three dimensions,
are not confined to a stationary plane, and are dynamically intricate.
A complete classification of these $3D$ orbits is the purview of this
article.

Carter famously showed that there were four constants of
motion\cite{carter1968, misner} for the orbits of spinning black
holes, one for each canonical momentum, so that the orbits are
integrable.  Still, black hole orbits have long evaded a simple
geometric classification.  While any geodesic orbit could be computed
easily, a concise general account of how changes to the constants of
motion would alter its shape was unavailable.  Recently a topological
taxonomy based on periodic orbits provided a complete classification
of all \emph{equatorial} orbits \cite{levin2008}.

In brief, Ref.\ \cite{levin2008} shows that just as Mercury is a
precession of the ellipse, any relativistic orbit can be understood as
a precession of a periodic orbit. Although there is no ellipse in
relativity, no 1-leaf clover, there are 2-leaf, 3-leaf,... $n$-leaf
clovers as well as $n$-leaf clovers with nearly circular whirls. The
equatorial periodic orbits are defined by a rational number
\begin{equation}
q_{r\varphi}=\frac{\omega_\varphi}{\omega_r}-1
\end{equation}
where $\omega_\varphi $ is an average angular frequency in the
equatorial plane and $\omega_r$ is the radial frequency. Aperiodic
orbits correspond to irrational ratios of frequency while periodic
orbits correspond to rational $q_{r\varphi}$.  The number
$q_{r\varphi}$ explicitly measures the degree of perihelion precession
beyond the ellipse as well as the topology of the orbit. The
$q_{r\varphi}=1/3$ orbit is a 3-leaf clover while the
$q_{r\varphi}=1+1/3$ orbit is a 3-leaf clover with 1 whirl per radial
cycle.  And, importantly, the $q_{r\varphi}=1/3+\epsilon$ orbit looks
like a $3$-leaf clover precessing at a rate of $2\pi\epsilon$ of
azimuth per radial cycle. (For a complete description see
Ref.\ \cite{levin2008,levin2009}.) The classification is especially
effective since $q_{r\varphi}$ varies monotonically with the energy of
an orbit for a given $L$.  As the value of $q_{r\varphi}$ increases,
the topology of the orbit varies in a systematic way as the energy and
orbital eccentricity also increase (for a given $L$).  The resulting
taxonomy nicely exposes the complete equatorial dynamics.

\begin{figure*}
  \includegraphics[scale=0.54]{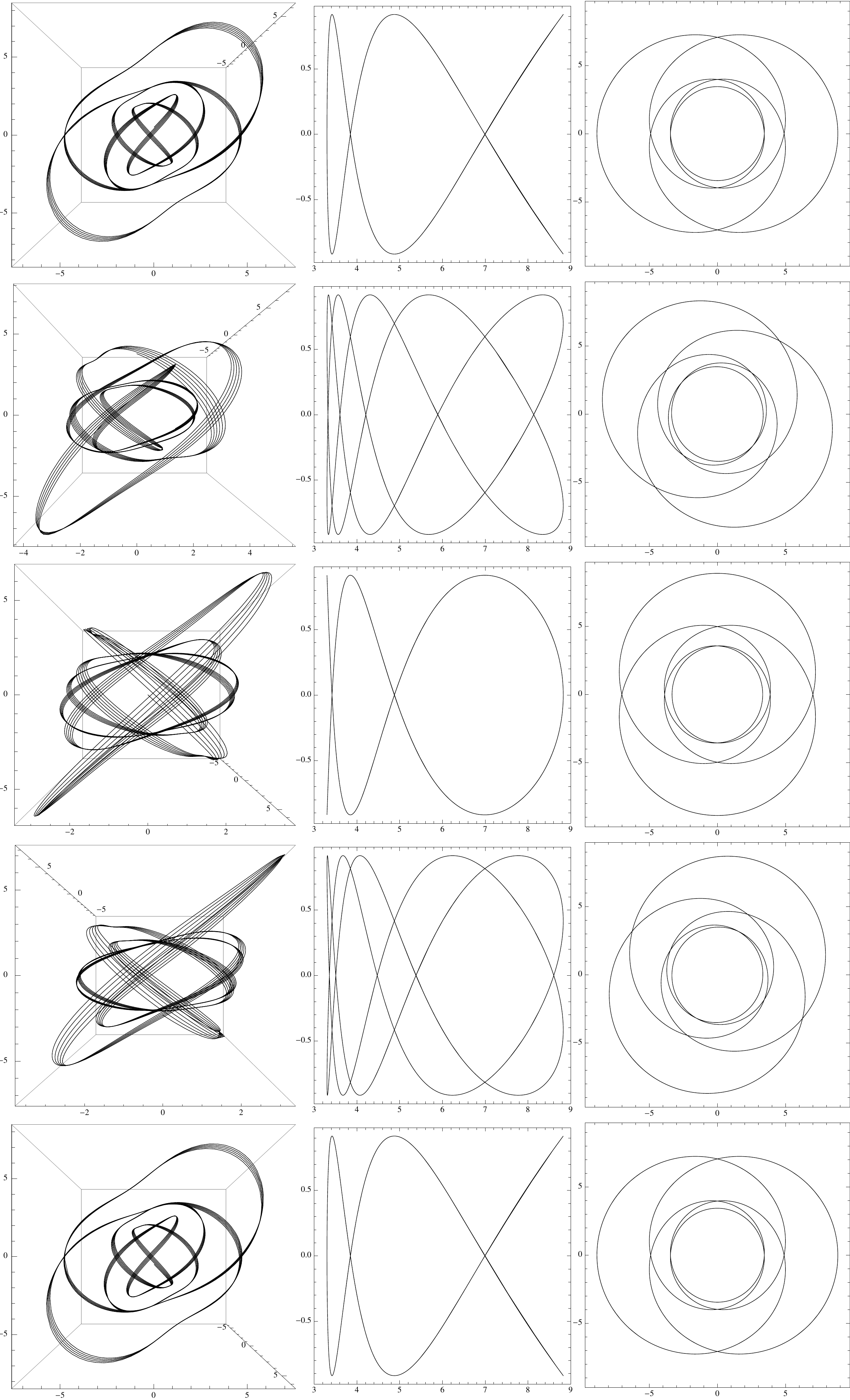}
  \caption{$r$-$\theta$ periodic orbits with $L=3$, $E=0.932516$,
    $\cos{\iota}=0.4$ and $a=0.99$, but with different $r-\theta$
    phasing.  Column 1 shows the full $3$D orbit. Column 2 is the
    projection of the orbit into the $r$-$\cos{\theta}$ plane.  Column
    3 is the projection into the orbital plane.  All rows have
    $r_{0}=r_{a}=8.82713$ and $\varphi_{0}=0$. The initial $\theta$
    values are as follows: Row 1 $\theta_{0}=\theta_{\rm
      min}=0.414139$; Row 2 $\theta_{0}=0.8$; Row 3
    $\theta_{0}=\frac{\pi}{2}$; Row 4 $\theta_{0}=2$; Row 5
    $\theta_{0}=\theta_{\rm max}=2.72745$.}
  \label{fig:per_int_cond}
\end{figure*}

The goal here is to generalize the equatorial taxonomy to fully
generic $3D$ Kerr motion. We could identify fully periodic orbits and
argue that all generic orbits are approximated at arbitrary precision
by that set of measure zero \cite{levin2008}. However, it is
sufficient to consider the less restrictive, larger set of orbits that
are periodic only in $r-\theta$, as these will be shown to be
perfectly periodic when projected into an instantaneous orbital plane,
as illustrated in Fig.\ \ref{fig:per_int_cond}.

A series of orbits is shown in $3D$ on the leftmost column of
Fig.\ \ref{fig:per_int_cond}, in the $r-\cos{\theta}$ plane in the
middle column, and projected in an effective orbital plane in the
final column. These orbits are closed in $r-\theta$ and also in the
orbital plane, but are not fully closed in $3D$. The following
sections will be devoted to realizing this argument.  Similar
reasoning led to a taxonomy of generic $3D$ orbits in a Post-Newtonian
expansion of two black holes in Ref.\ \cite{levin2008:2,grossman2008}.

\section{The Basics}
\label{sec:thebasics}

We begin with the Kerr metric in Boyer-Lindquist coordinates and
geometrized units $\left(G=c=1\right)$ and the conventional choice of
$M=\mu=1$:
\begin{align}
\label{eq:kerr_metric}
ds^{2} = &-d{\tau}^2 \\ \nonumber =
&-\left(1-\frac{2r}{\Sigma}\right)dt^{2}
-\frac{4ar\sin^{2}{\theta}}{\Sigma}dtd\varphi+
\frac{\Sigma}{\Delta}dr^{2} + \Sigma d{\theta}^2 \\ \nonumber &+
\sin^{2}{\theta}\left(r^{2} + a^{2} +
\frac{2a^{2}r\sin^{2}{\theta}}{\Sigma}\right)d{\varphi}^2 \quad ,
\end{align}
where
\begin{align}
\label{eq:sigma_delta}
\Sigma &\equiv r^{2} + a^{2}\cos^{2}{\theta}
\\
\nonumber
\Delta &\equiv r^{2} - 2r +a^{2}
\quad .
\end{align}
Carter reduced the equations to first integrals of
motion \cite{carter1968, misner}, exploiting the four constants of motion
$E,L_z,Q$ and $\mu$:
\begin{subequations}
\label{eq:dimcarter}
\begin{alignat}{1}
\dot r &= \pm \sqrt{R}
\label{subeq:dimcarter-r}\\
\dot \theta &= \pm \sqrt{\Theta}
\label{subeq:dimcarter-theta}\\
\dot \varphi &=
\frac{a}{\Delta} \left( 2rE - aL_z \right)
+ \frac{L_z}{\sin^2 \theta}
\qquad \quad ,
\label{subeq:dimcarter-phi}\\
\dot t&=
\frac{(r^2 + a^2)^2 E - 2arL_z}{\Delta}
- a^2 E \sin^2 \theta \quad\quad .
\label{subeq:dimcarter-t}
\end{alignat}
\end{subequations}
We will often refer to equations (\ref{eq:dimcarter}) as the Carter
equations.  In those equations, an overdot denotes differentiation
with respect to Mino time \cite{mino2003}, $\lambda$, which is related
to the particle's proper time, $\tau$, by $d\lambda =
\frac{d\tau}{\Sigma}$, and the quantities
\begin{alignat}{1}
  \Theta(\theta) &= Q - \cos^2\theta
  \left\{
  a^2(1- E^2) + \frac{L_z^2}{\sin^2\theta}
  \right\}
  \label{eq:Thetaeq}\\
  \begin{split}
    R(r) &= -(1 - E^2)r^4 + 2r^3 - \left[ a^2(1 - E^2) + L_z^2 \right]r^2 \\
    &=  {}+ 2(aE - L_z)^2\, r - Q \Delta
  \end{split}
  \label{eq:Rpoly}
\end{alignat}
are the polar and radial quasi-potentials, respectively \cite{wilkins1972}.

The quasi-potentials reveal some well-known geometric information
about bound non-plunging orbits (orbits that neither escape to
infinity nor cross the horizon of the central black hole).  First,
they reveal the radial turning points, which occur at roots of $R(r)$.
For a given $E, L_{z}$ and $Q$, the quartic polynomial has four roots.
The outermost two are periastron and apastron,
between which the radial position of a bound orbit oscillates.  
A
similar analysis of the roots of $\Theta(\theta)$ reveals that every
bound orbit oscillates between a fixed $\theta_{\rm min}$ and
$\theta_{\rm max}$ symmetrically distributed about the equatorial
plane\footnote{For equatorial orbits, $\theta_{\text{min}}
  =\theta_{\text{max}} \equiv \pi/2$.}, i.e.\ $\theta_{\rm
  min}=\pi-\theta_{\rm max}$
\cite{levin2008:3,wilkins1972,hughes2001}.  The upshot is that every
$3D$ orbit will generally lie in a toroidal wedge around the
equatorial plane bounded $r_{p}$ and $r_{a}$ in radial coordinate and
between $\theta_{\rm max}$ and $\pi-\theta_{\rm max}$ in polar angle
\cite{drasco2006}.

Every bound Kerr orbit also has an associated triplet of fundamental
frequencies $(\omega_{r}, \omega_{\theta}, \omega_{\varphi})$ , which
can be defined for any choice of time coordinate \cite{schmidt2002}.
The simplicity afforded by the choice of Mino time and exploited
heavily in \cite{mino2003, drasco2004} is that, since the radial and
polar motions decouple in Mino time, each of $r(\lambda)$ and
$\theta(\lambda)$ is independently periodic.  As a result, the
Mino-time frequencies can be defined and computed directly from
equations (\ref{subeq:dimcarter-r}) and (\ref{subeq:dimcarter-theta}).

We will only be concerned with the radial and polar frequencies here.
To obtain them, we first define the radial and polar Mino periods via
\begin{subequations}
\begin{alignat}{1}
\Lambda_r=2\int_{r_p}^{r_a}\frac{d\lambda}{dr}{dr} &=
2\int_{r_p}^{r_a}\frac{ dr}{\sqrt{R(r)}}
\\
\Lambda_\theta=4\int^{\pi/2}_{\theta_{\rm min}}
\frac{d\lambda}{d\theta}d\theta &=
4\int^{\pi/2}_{\theta_{\rm min}} \frac{d\theta}{\sqrt{\Theta(\theta)}}
\quad .
\end{alignat}
\end{subequations}
The corresponding Mino-time frequencies are then
\begin{subequations}
\begin{alignat}{1}
  \omega_{r} &\equiv \frac{2\pi}{\Lambda_{r}}
  \\
  \omega_{\theta} &\equiv \frac{2\pi}{\Lambda_{\theta}}
\quad .
\end{alignat}
\end{subequations}
Note that we use Mino time purely for ease and convenience and that
the frequency \emph{ratios} which figure prominently in our analysis
are independent of the choice of time variable.

We want to consider $3D$ orbits that are closed in $r-\theta$.  That
closure will result when the polar and radial frequencies are
rationally related, or in language more directly useful for our
orbital plane description of the motion, when the quantity
\begin{equation}
q_{r\theta} \equiv \frac{\omega_\theta}{\omega_r}-1
\end{equation}
is rational.  To be useful, a classification based on orbits with
rational $q_{r\theta}$ has to have two properties: the rational
$q_{r\theta}$ must tell us about the topology of the orbit, and it must
relate that topology to more physical conserved quantities.  In the
subsequent sections, we show that this is indeed the case.

\subsection{The Energy Spectrum}

\begin{figure*}
  \includegraphics[scale=0.75]{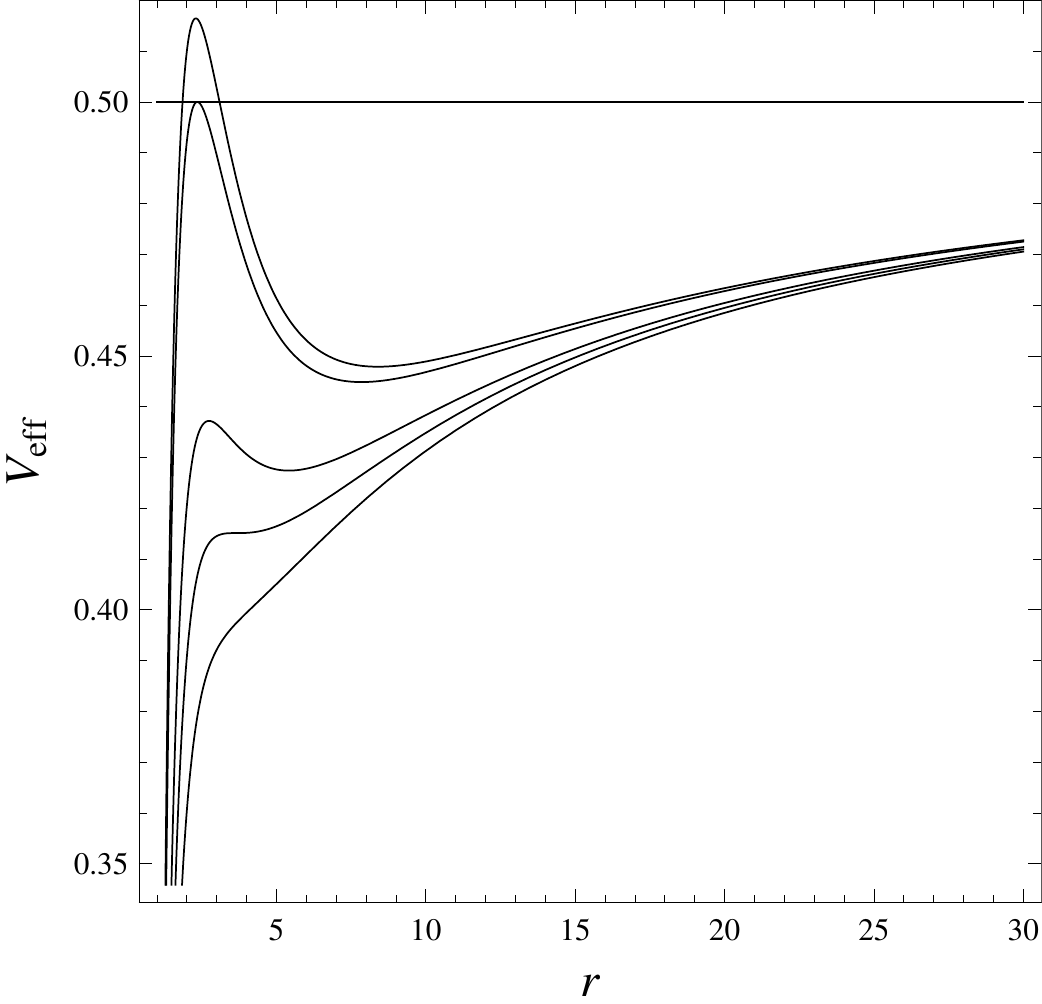}
  \includegraphics[scale=0.75]{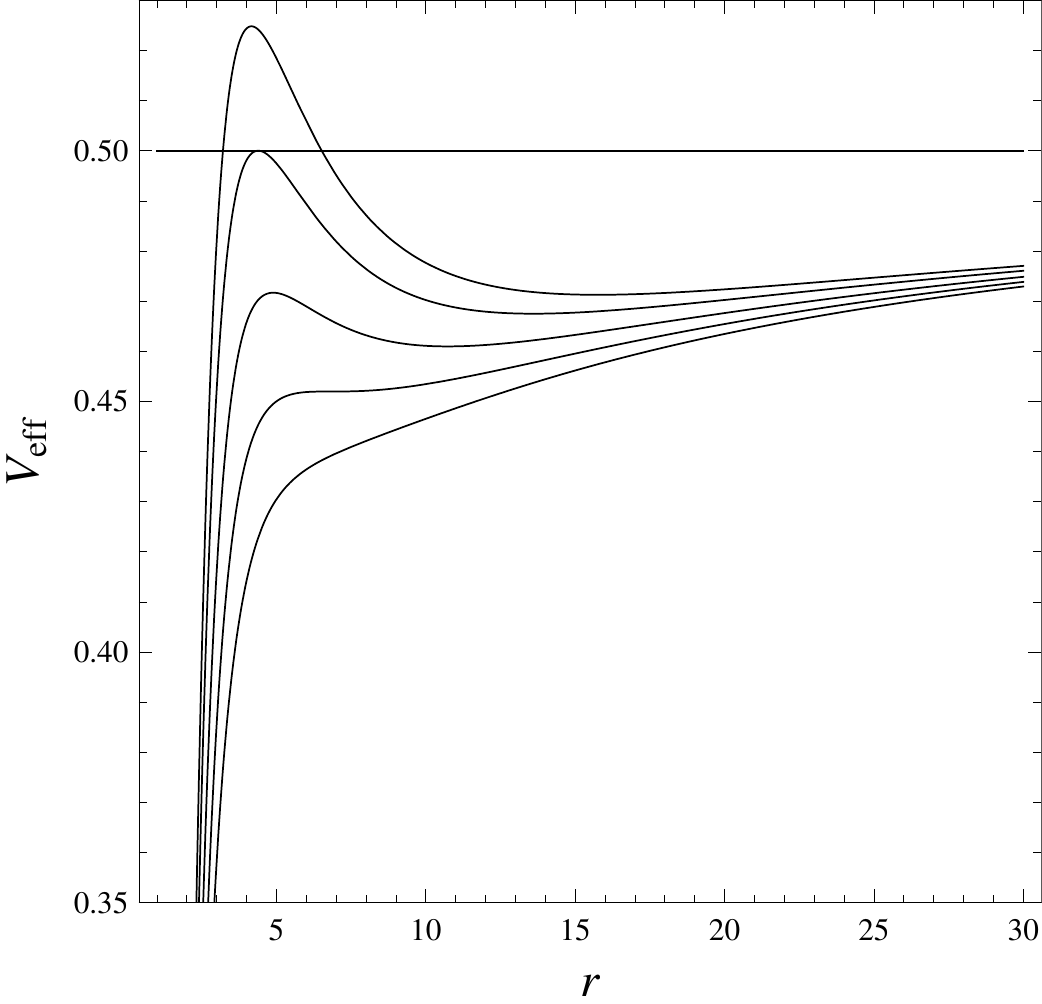}
  \caption{Left: Five $V_{\rm eff}$ curves with different $L$ values
    but all with $a=0.99$ and $\cos{\iota}=0.4$.  In order from the
    highest to the lowest curve, the $L$ values are: $L=3.4 >
    L_{\text{ibso}}$, $L=L_{\text{ibso}}=3.32432$,
    $L_{isso}<L=3<L_{\text{ibso}}$, $L=L_{\text{isso}}=2.85501$, and
    $L=2.7<L_{\text{isso}}$.  The horizontal line at $V_{\rm
      eff}=\frac{1}{2}$ shows the energy of marginolly bound orbits.
    Right: Five $V_{\rm eff}$ curves with different $L$ values but all
    with $a=0.99$ and $\cos{\iota}=-0.4$.  The highest curve has an
    $L=4.5 > L_{\text{ibso}}$.  The second highest curve has
    $L=L_{\text{ibso}}=4.28330$.  The middle curve has
    $L_{\text{isso}}<L=4<L_{\text{ibso}}$.  The second shortest curve
    has an $L=L_{\text{isso}}=3.74594$.  And the smallest curve has
    $L=3.5<L_{\text{isso}}$.  The horizontal line at $V_{\rm
      eff}=\frac{1}{2}$ shows the energy of the marginolly bound
    orbits.}
  \label{fig:Veff_kerr_general}
\end{figure*}

In the spirit of the equatorial classification of \cite{levin2008}, we
begin by describing how $q_{r\theta}$ varies with energy.  For ease,
and in anticipation of the fact that our analysis will ultimately
focus on the discrete set of $E$ values for those orbits with rational
values of $q_{r\theta}$, we will refer (loosely) to the dependence of
$E$ on $q_{r\theta}$ as an energy spectrum.  The subtlety in
establishing a simple relationship between $q_{r\theta}$ and $E$ is
the choice of which other parameters to keep fixed as $E$ is varied.
In the appendix we show that the key combinations are an effective
total angular momentum $L$ and angle of inclination $\iota$ for orbits
around a black hole of a given spin $a$ defined by
\begin{align}
L^{2} &\equiv L_{z}^{2} + Q
\\
\nonumber
\cos{\iota} &\equiv \frac{L_{z}}{L}
\quad ,
\end{align}
first used by \cite{ryan1995, ryan1996} and used occasionally in other
references \cite{drasco2006,hughes2001,hughes2001:2,glampedakis2002:2}.  Our
construction turns out to be greatly facilitated by varying $L$ while
keeping $\iota$ fixed, as opposed to varying $L_z$ while keeping $Q$
fixed.

This choice of orbital parameters allows us to write equations
(\ref{eq:Thetaeq}) and (\ref{eq:Rpoly}) as
\begin{alignat}{1}
  \Theta(\theta) &= L^{2}\sin^{2}{\iota} - \cos^2\theta
  \left\{
  a^2(1- E^2) + \frac{L^2\cos^{2}{\iota}}{\sin^2\theta}
  \right\}
  \label{eq:Thetaeq_iL}\\
  \begin{split}
R\left(r\right) &= \left(E^2 - 1 \right) r^4 + 2r^3 + \left(a^2
  \left\{E^2 - 1\right\} - L^2 \right)r^2 \\
  &+ 2r \left(a^2 E^2 - 2 a E L\cos{\iota}
  + L^2 \right) \\
&+ a^{2}L^{2}\left(\cos^{2}{\iota} -1 \right)
    \qquad .
  \end{split}
  \label{eq:Rpoly_iL}
\end{alignat}
With this particular combination of constants, we can produce an
analog of the familiar Schwarzschild effective potential for
nonequatorial Kerr motion.  Consider a black hole of given spin
$a$. For a non-spinning black hole ($a=0$), we can rewrite the radial
equation (\ref{subeq:dimcarter-r}) as
\begin{equation}
\label{eq:Schwarz_Veff_eqn}
\frac{1}{2}\left (\frac{dr}{d\tau}\right )^2+V_{\rm eff}=\epsilon_{\rm eff}
\quad .
\end{equation}
This standard effective potential formulation of Schwarzschild motion
relates the radial velocity with respect to particle proper time to an
effective energy $\epsilon_{\rm eff}=E^2/2$ and an effective
potential $V_{\rm eff}$ that is a different function of $r$ for each
fixed $L$ -- crucially, $V_{\text{eff}}$ is independent of $E$.  The
result is a simple visual way to describe the different types of
allowed motion as $L$ is varied.

However, for fully $3D$ orbits around a spinning black hole, an
analogous potential is not self-evident.  The counterpart to equation
(\ref{eq:Schwarz_Veff_eqn}) (which must involve velocities with
respect to Mino-time in order to decouple the radial motion from the
polar motion) is
\begin{equation}
\label{eq:rearrange}
\frac{1}{2}\dot r^2- \frac{R}{2}=0
\quad .
\end{equation}
In this case, the dependence on $E$ in $R(r)$ cannot be simply
separated and moved to the right-hand side.  It would seem that the
best we can do with eqn.\ (\ref{eq:rearrange}) is end up with a
$V_{\rm eff}$ that depends on \emph{all} the constants of motion.  We
therefore lose the ability to visualize easily the variation of orbits with
energy, as we can in the Schwarzschild case, because even at fixed
$L_{z}$ and $Q$ (or $L$ and $\iota$), changing $E$ also causes the potential to shift.  As written, then, equation
(\ref{eq:rearrange}) admits a one-dimensional effective potential
description, but that description is not \emph{useful} because there
is a different potential for every combination of orbital parameters.

However, if we consider only the behavior at the {\it turning points},
we can construct a useful pseudo-effective potential.  The idea is to
set $\dot r=0$ in Eqn.\ (\ref{eq:rearrange}), which amounts to setting
$R(r)=0$, and to solve for $E$:
\begin{widetext}
\begin{align}
    \label{eq:V_eff}
E\left(a,\iota, L, r \right) &=
\frac{2a\cos{\iota}Lr+\sqrt{r\left(a^{2}+\left(-2+r\right)r\right)\left(r^{3}\left(L^{2}+r^{2}\right) +a^{2}\left(2+r\right)\left(L^{2}-\cos^{2}{\iota}L^{2}+r^{2}\right)\right)}}{r\left(r^{3}+a^{2}\left(2+r\right)\right)}
\end{align}
\quad .
\end{widetext}   
We then define a pseudo-effective potential
\begin{equation}
\left. V_{\rm eff}\right |_{\dot r=0}=\frac{E^2}{2} \quad .
\end{equation}
that will allow us to draw various lines of fixed $E$ on a potential
that maintains its shape and visually identify turning points of the
motion, even if the difference between $E$ and the value of
$V_{\rm eff}$ no longer gives the value of $\dot{r}^2$.

Fig.\ \ref{fig:Veff_kerr_general} illustrates the utility of this
approach. For every fixed $\iota$, as $L$ is lowered we get an analogous
pattern of orbits to the Schwarzschild case. There is a minimum of
$V_{\rm eff}$, which corresponds to a stable constant radius orbit. There
is a maximum of $V_{\rm eff}$, which corresponds to an unstable constant
radius orbit. Unlike the Schwarzschild case, the constant radius
orbits are not circles. Instead they lie on the surface of a sphere
bounded between $\theta_{\rm max}$ and $\theta_{\rm min}=\pi-\theta_{\rm max}$. We hereafter call
these spherical orbits. 

The parallel story continues.  As $L$ is lowered, the unstable
spherical orbit becomes bound once a certain critical $L$ value is
crossed.  That value $L_{\text{ibso}}$ is the angular momentum of the
innermost bound spherical orbit (ibso), the unstable spherical orbit
with critical energy $E=1$.  An innermost stable spherical orbit
(isso) appears as a saddle point of $V_{\text{eff}}$ once $L$ drops to
yet another critical value $L_{\text{isso}}$.  For
$L<L_{\text{isso}}$, all orbits plunge into the central black hole.
Fig.\ \ref{fig:Veff_kerr_general} demonstrates the consistency for
both prograde and retrograde orbits.

If we had chosen to keep $L_z$ fixed while varying $Q$ instead of
keeping $L$ fixed while varying $\iota$, we would not have seen the
same simple pattern. Appendix \ref{sec:cons_quant} shows the breakdown
in the Schwarzschild analogy when using orbital parameters
$(L_{z},Q)$.

The result, for a given $L,\iota$, is that $q_{r\theta}$ increases
monotonically with energy.  The lowest energy bound orbit is the
stable spherical orbit, and, importantly, this orbit has the lowest
value of $q_{r\theta}$ for that combination of $L,\iota$. As detailed
in Ref.\ \cite{levin2008}, the constant radius orbits do not have
rational value zero, as can be proven by taking the zero eccentricity
limit, $e\rightarrow 0$.

Since $q_{r\theta}$ is monotonic, its upper bound
$q_{r\theta}^{\text{max}}$ is the value of $q_{r\theta}$ for the
maximum energy bound non-plunging orbit for a given $L$.  Whether
$q_{r\theta}^{\text{max}}$ is finite or infinite depends on whether
$L$ is greater than or less than $L_{\text{ibso}}$.  If
$L>L_{\text{ibso}}$, the unstable spherical orbit is unbound and has
energy $E > 1$.  $q_{r\theta}^{\text{max}}$ is therefore the
$q_{r\theta}$ value of the $E=1$ orbit, and despite the fact that the
$E=1$ orbit just reaches $r=\infty$ after infinite time, its
$q_{r\theta}$ is nonetheless finite.  As we reduce $L$,
$q_{r\theta}^{\text{max}}$ increases monotonically, and eventually
$q_{r{\theta}}^{\text{max}}\rightarrow \infty$ once
$L=L_{\text{ibso}}$. For all $L<L_{\text{ibso}}$,
$q_{r{\theta}}^{\text{max}}$ remains infinite \cite{{levin2008:3},
  {perez-giz2008}}. This happens because the maximum energy bound
non-plunging orbit is now the homoclinic orbit (or separatrix orbit),
which formally has an infinite number of whirls during its lone
infinite-period radial cycle. A detailed analysis of the homoclinic
orbit can be found in \cite{{levin2008:3}, {perez-giz2008}}.

\begin{figure}
\center
  \includegraphics[scale=0.75]{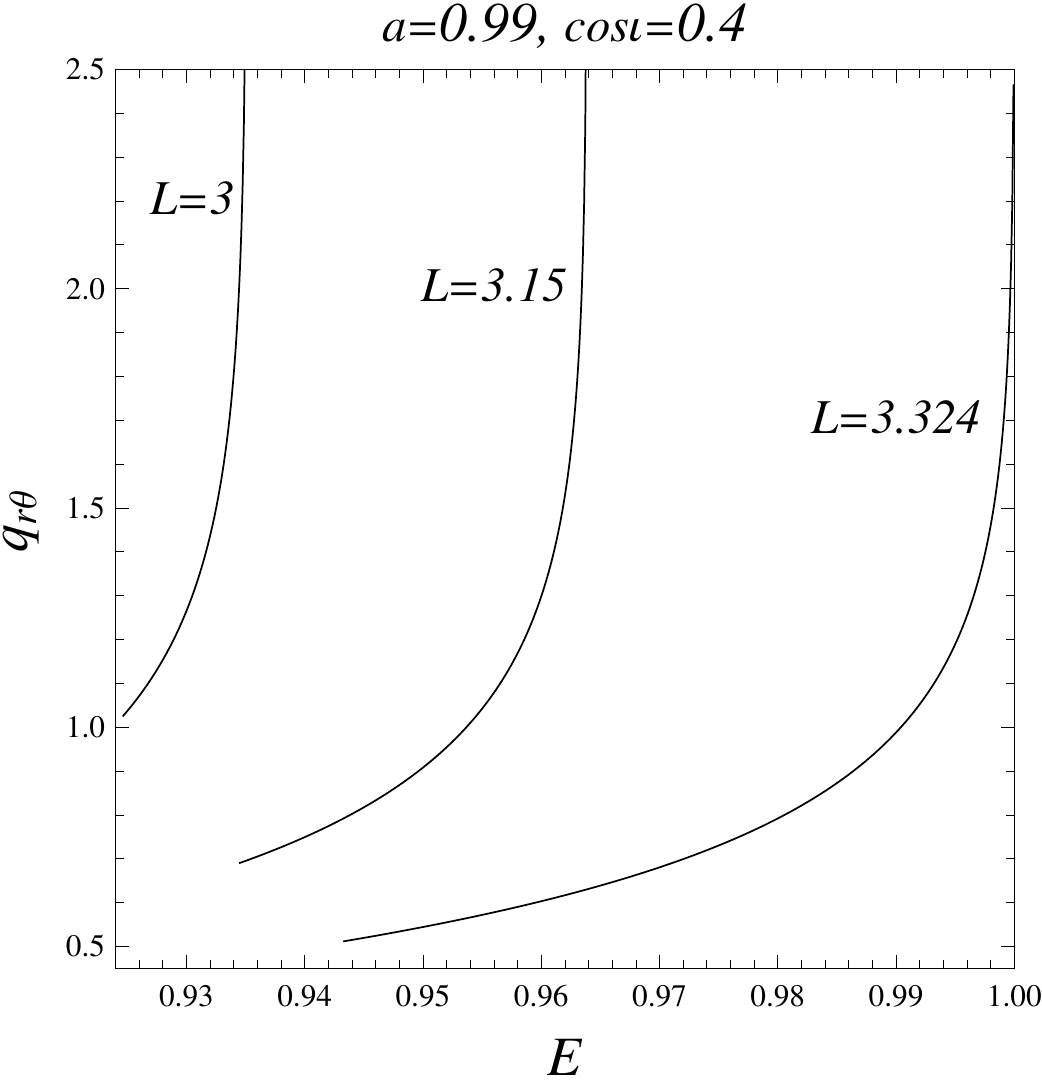}
  \includegraphics[scale=0.75]{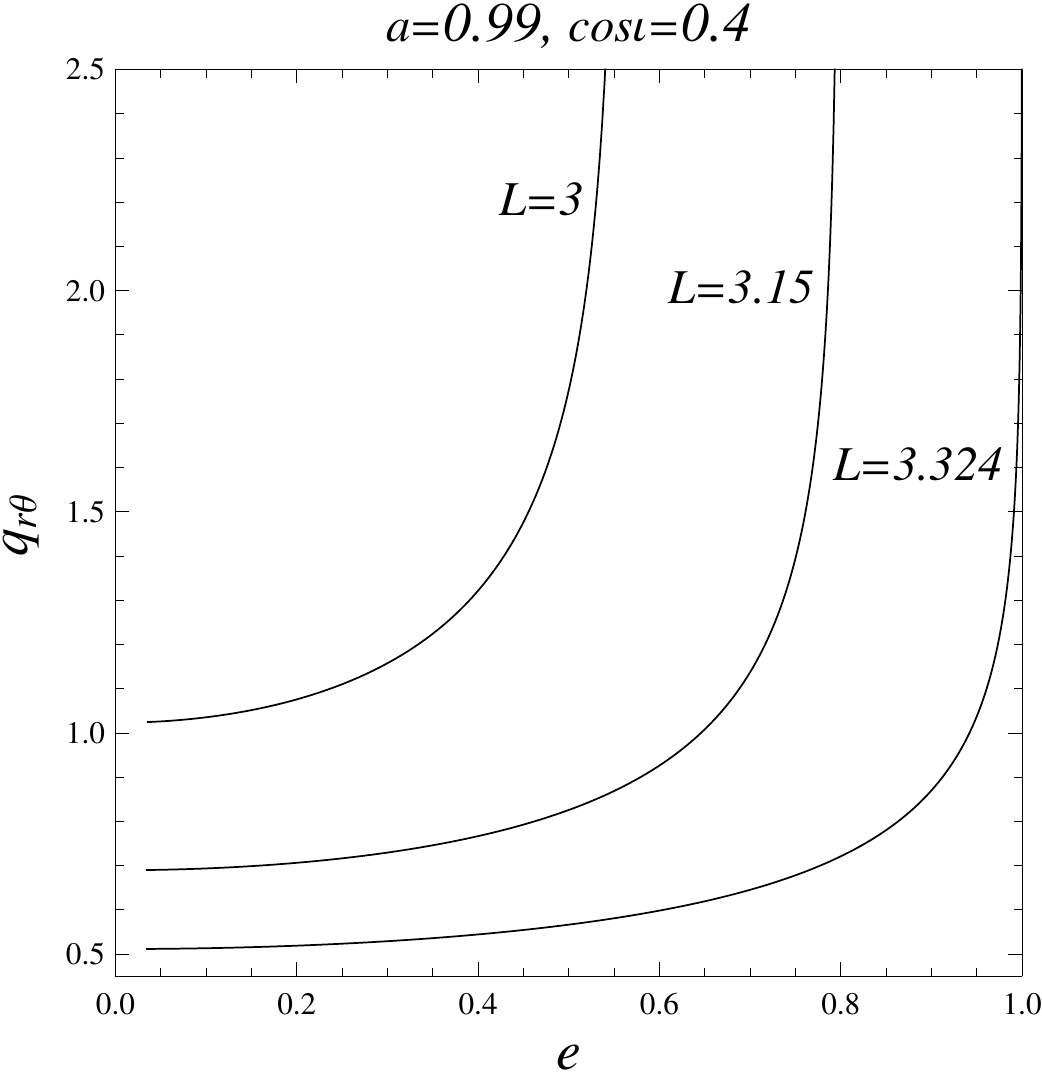}
  \caption{Top: The plot shows the monotonic relationship between
    $q_{r\theta}$ and energy for all bound orbits with a given $a$,
    $L$ and $\cos{\iota}$. We show three different $L$ values all with
    $a=0.99$ and $\cos{\iota}=0.4$. The graphs cut off on the left at
    the energy value for the stable spherical orbit with that $a,
    \iota$ and $L$.  Bottom: The plot shows, for the above parameter
    values, the monotonic relationship between $q_{r\theta}$ and
    orbital eccentricity $e \equiv \frac{r_a - r_p}{r_a +r_p}$.  The
    lower eccentricity bound is $e=0$, also corresponding to the
    stable spherical orbits.}
  \label{fig:E_q_rth_mono}
\end{figure}

Figure \ref{fig:E_q_rth_mono} is a plot of the $q_{r\theta}$ versus
energy for a given $a,\iota$ and $3$ sets of $L$ values.  It is
representative of the general trend we see for any combination\footnote{The case of $\iota=0,\pi$ needs to be handled as
  in Ref. \cite{levin2008} because that is motion that takes place
  entirely in the equatorial plane.}  of
$a,L,\iota$. As the energy increases, so does
$q_{r\theta}$.  As $L$ decreases towards $L_{\text{isso}}$, the minium value
of $q_{r\theta}$ increases.  This trend was seen equatorially in
Ref.\ \cite{levin2008}.

In figure \ref{fig:E_q_rth_mono} we see that the $q_{r\theta}$ also
increases with eccentricity, $e$.  Again this is a general trend so
that $q_{r\theta}$ is monotonic with eccentricity. The larger
$q_{r\theta}$, again for a fixed $(a,L,\iota)$, the more eccentric the
orbit.  

We have shown that $q_{r\theta}$ corresponds to an energy spectrum for
$3D$ orbits. What we want now is to show this also corresponds to a
measure of zoom-whirliness and so is also a toplogical indicator.  As
we will see, quite incredibly, this $q_{r\theta}$ measures the amount
by which the angle {\it in the orbital plane} overshoots $2\pi$, that
is, precesses, in one radial period. In other words, when
$q_{r\theta}$ is rational, it is a direct measure of the topology of
the orbit in the orbital plane and increases monotonically with
energy, thereby defining a spectrum of zoom-whirl orbits in the
orbital plane.

\subsection{Periodic Tables and the Orbital Plane}

\begin{figure*}
  \centering
  \includegraphics[scale=0.65]{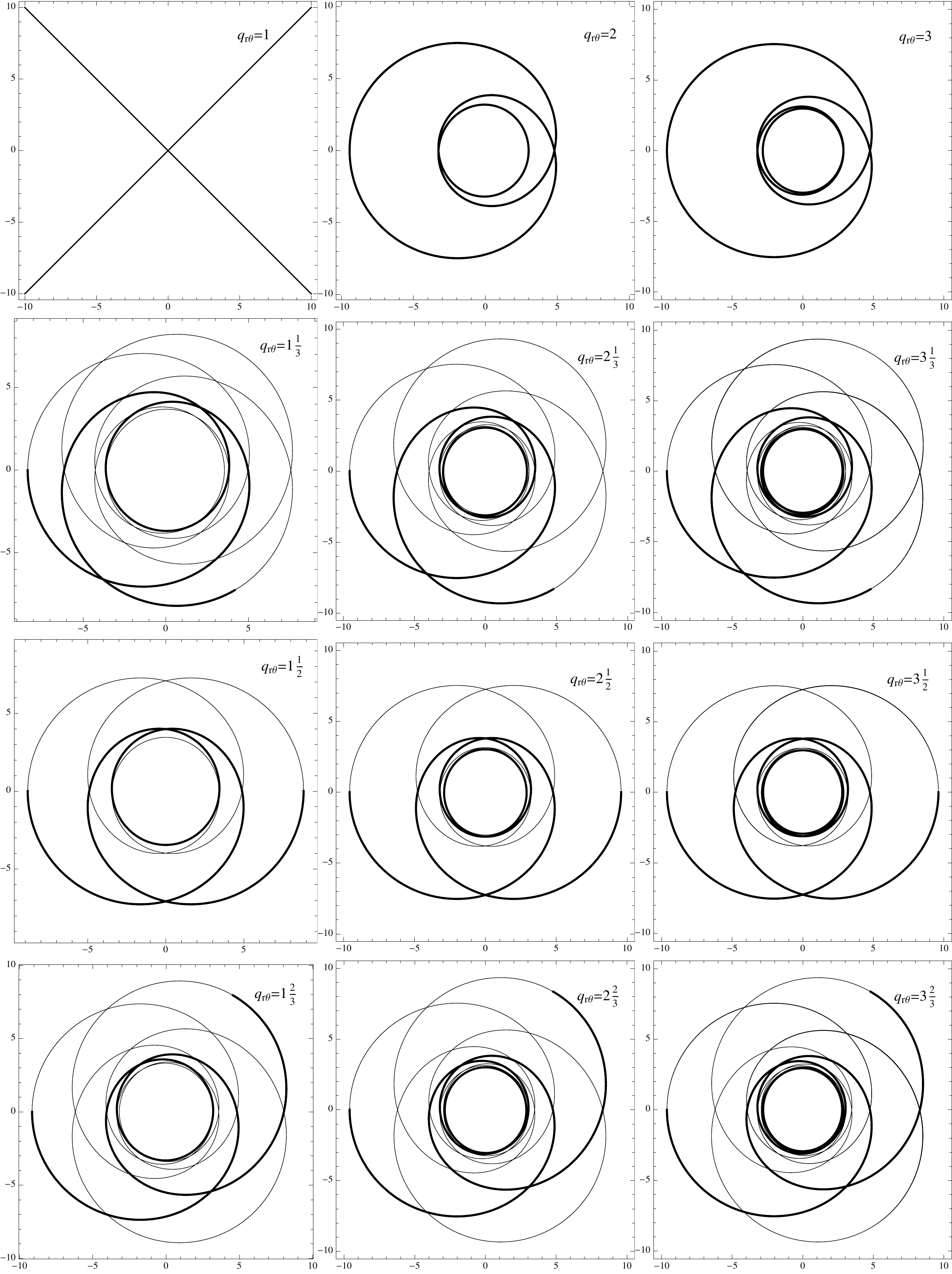}
  \caption{A periodic table for which the orbits have been progected
    into the orbital plane.  All orbits were started at $r_{0}=r_{a}$
    and $\theta_{0}=\theta_{\rm max}$.  The orbital parameters are:
    $a=0.99$, $L=3$, $\cos{\iota}=0.4$.  The energy increases from top
    to bottom and left to right.}
  \label{fig:orb_plane_periodic}
\end{figure*}

\begin{figure*}
  \centering
  \includegraphics[scale=0.65]{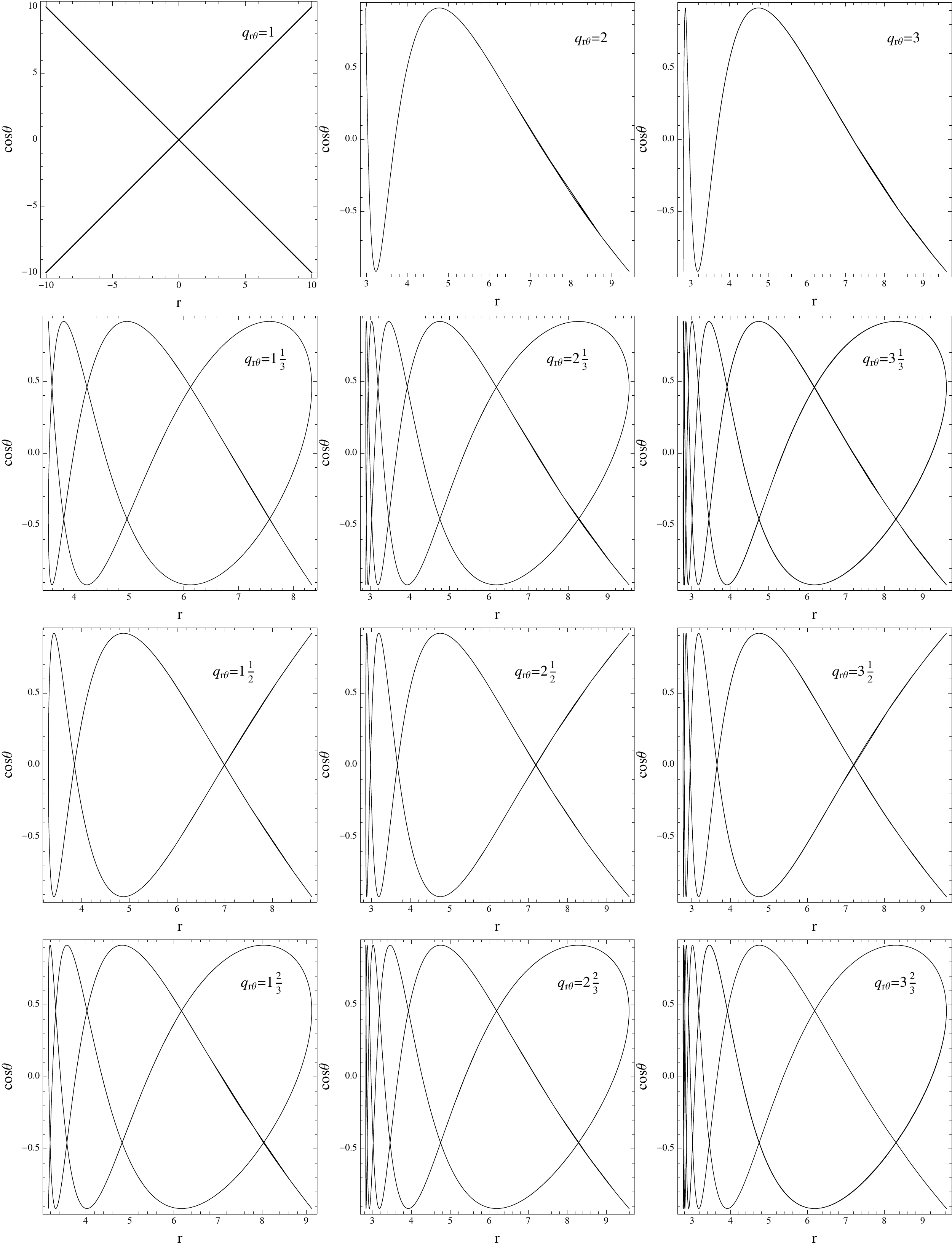}
  \caption{A periodic table for which the orbits have been progected
    into the $r$-$\cos{\theta}$ plane.  All orbits were started at
    $r_{0}=r_{a}$ and $\theta_{0}=\theta_{\rm max}$.  The orbital
    parameters are: $a=0.99$, $L=3$, $\cos{\iota}=0.4$.  The energy
    increases from top to bottom and left to right.}
  \label{fig:liss_periodic}
\end{figure*}
 
We preface this section with the caveat that the orbital plane
construction below naively employs flat space vector algebra and
vector calculus constructions (e.g.\ cross products of 3-vectors)
without fully taking into account the curvature of the background Kerr
spacetime.  Prima facie, it is not obvious that the formalism should
accurately capture geometric or topological features of 3D orbits.
Nevertheless, we have the amazing result that the $r-\theta$ periodic
orbits correspond to a spectrum of zoom-whirl orbits in this effective
orbital plane, beautifully mirroring the equatorial result of
Ref.\ \cite{levin2008}.  For now, we simply state our results, which
are compelling, and report that a more precise analysis of the
connection between the orbital plane construction and a
relativistically precise projection of the motion using local tetrads
is underway. A very precise implementation for the PN-expansion of two
black holes can be found in Refs.\ \cite{levin2008:2,grossman2008}.

We consider the projection of $r-\theta$ periodic orbits in an
instantaneous orbital plane that we define naively as the plane in the
tangent space spanned by $\vec R$ and $\vec P$, defined below, with a corresponding
angular momentum $\vec \Lo=\vec R\times \vec P$.  At every instant, the
orbital plane is the plane perpendicular to the angular momentum
vector.  

It is useful to define
\begin{align}
\rho= &(r^2+a^2)^{1/2}
\end{align}
and convert from ellipsoidal to Cartesian coordinates
\begin{align}
x=&\rho\sin\theta\cos\varphi \nonumber \\
y=&\rho\sin\theta\sin\varphi \nonumber \\
z=& r\cos\theta  \quad .
\end{align}
Then,
\begin{equation}
\vec \Lo=\vec R\times \vec P
\end{equation}
where
\begin{align}
\vec R& =(x,y,z) \nonumber \\
\vec P & =(P_x,P_y,P_z)\nonumber 
\end{align}
for which
\begin{equation}
P^i=\frac{\partial x^i}{\partial q^j}g^{kj} P_k 
\end{equation}
where $i=x,y,z$ and $k,j=r,\theta,\varphi$.
For convenience we take
the $M\rightarrow 0$ limit \cite{carroll},
\begin{eqnarray}
ds^2 &=& -dt^2+\frac{(r^2+a^2\cos^2\theta)}{(r^2+a^2)}dr^2
\\
\nonumber
& & +(r^2+a^2\cos^2\theta)d\theta^2+ (r^2+a^2)\sin^2\theta d\varphi^2
\end{eqnarray}
so that
\begin{align}
P_x =&\frac{r\rho}{{\Sigma}}\sin\theta\cos\varphi P_r+\frac{\rho}{\Sigma}\cos\theta\cos\varphi P_\theta
    -\frac{\sin\varphi}{\rho\sin\theta}P_\varphi\nonumber \\
P_y =&\frac{r\rho}{{\Sigma}}\sin\theta\sin\varphi P_r+\frac{\rho}{\Sigma}\cos\theta\sin\varphi P_\theta
    +\frac{\cos\varphi}{\rho\sin\theta}P_\varphi \nonumber \\
P_z = & \frac{\rho^2}{{\Sigma}}\cos\theta P_r
-\frac{r}{\Sigma}\sin\theta P_\theta \quad .
\end{align}
To find the orbital plane, we write
\begin{align}
\vec \Lo=&\Lo_z\hat k+\Lo_{\perp}\hat \perp \nonumber \\
\Lo_{\perp}\hat \perp= &\Lo_x\hat i+\Lo_y\hat j 
\end{align}
so that we can define
\begin{align}
\hat X=& \hat k\times \hat \perp \nonumber \\
\hat Y =& \hat \Lo\times \hat X \quad .
\end{align}
The orbital plane is spanned by $\hat X, \hat Y$. (For a more detailed
exposition on the orbital plane variables, see
Ref.\ \cite{levin2008:2,grossman2008}.)  This informally defined
orbital plane is sufficient, as we will see, since it effectively
soaks out any $\varphi$ motion.

Fig.\ \ref{fig:orb_plane_periodic} shows a table of orbits in the
effective orbital plane.  Our periodic table assembles orbits with
rational $q_{r\theta}$ as an energy spectrum, with energy increasing
from top to bottom and then from left to right.  The topology of
zoom-whirl orbits in the effective orbital plane is encoded in
$q_{r\theta}$ through
\begin{equation}
q_{r\theta}=w+\frac{v}{z}
\quad ,
\end{equation}
where $w$ is the number of nearly circular whirls and $v$ indicates
the order in which the $z$ zooms, or leaves, are traced out. So the
$q_{r\theta}=1+2/3$ orbit is a $(z=3)$-leaf clover, that executes
$w=1$ whirls during each each radial cycle before it moves to the
$v=2$ leaf in the pattern.

This result is quite remarkable: $q_{r\theta}$ is a measure of the
number of times the orbit returns to $\theta_{\rm min}$ per radial
cycle, yet it gives topological information about the degree of
precession in a very different angular variable, namely the angle
swept out in the oribtal plane.  Had we instead projected the orbit
onto the $r-\cos\theta$ plane, our $r-\theta$ periodic orbits would
look like Lissajous figures as in Fig.\ \ref{fig:liss_periodic}.  The
geometric information in Fig.\ \ref{fig:orb_plane_periodic} is
severely obscured when the trajectories are plotted as Lissajous
figures.

Fig.\ \ref{fig:per_int_cond} shows trajectories with the same orbital
parameters but different $r-\theta$ phasing.  All orbits
have the same $E,L,\iota$ and therefore the same
$(r_a,r_p,\theta_{\rm max})$. However,
$r_{a}$ coincides with different initial
values of $\theta$ in the range
$\pi-\theta_{\rm max}<\theta_o<\theta_{\rm max}$ for each picture.  Under shifts 
in $r-\theta$ phase, the $3D$ orbits
are all rather different (illustrated in
the first column) as are their corresponding Lissajous figures
(illustrated in the second column).  Notice, in stark contrast, that
varying the initial phasing of $r$-vs.-$\theta$ merely corresponds to
an overall rotation of the \emph{very same} zoom-whirl orbit in the
orbital plane (illustrated in the final column).

\section{Summary}

Our results are neatly summarized in Figures 
\ref{fig:orb_plane_periodic} and \ref{fig:per_int_cond}. 
Fig.\ \ref{fig:orb_plane_periodic} illustrates that
orbits periodic in
$r-\theta$ assemble into a spectrum of multi-leaf clovers
when projected in a loosely defined orbital plane. The
topology of the orbit is encoded in a rational number $q_{r\theta}=\frac{\omega_\theta}{\omega_r}-1$,
from which one can immediately read off the number of leaves (or zooms), the
ordering of the leaves, and the number of whirls. For a given
$L,\iota$, the rational number $q_{r\theta}$ monotonically increases
with
energy and with eccentricity. So, a simple $3$-leaf clover
($q_{r\theta}=1/3$) has less energy
and is less eccentric than a $2$-leaf ($q_{r\theta}=1/2$) of the same $L,\iota$.
Significantly, the rational number $q_{r\theta}$ is bounded below so
that there are no $q_{r\theta}\rightarrow 0$ orbits in the
strong-field regime. {\it There are therefore no tightly precessing
  elliptical orbits} in the strong-field regime. All eccentric orbits 
will have a countable number of leaves.

Moreover, as Fig.\ \ref{fig:per_int_cond} illustrates, a change in
$r-\theta$ phase corresponds to a simple rotation of the orbit in the
effective orbital plane. An orbit that hits apastron at $\theta_{\rm
  max}$ will be rotated by $\pi/2$ in the orbital plane relative to an
orbit with identical $(E,L,\iota)$ that hits apastron at $\theta=\pi/2$.

Any aperiodic orbit will be arbitrarily well-approximated by a nearby
periodic orbit. What's more, aperiodic orbits will look like
precessions of low-leaf clovers. Just as Mercury is a precession of
the ellipse, an orbit with $q_{r\theta}=1/2+\epsilon$ is the
precession of a $2$-leaf clover that accumulates an extra
$2\pi\epsilon$ of azimuth during each radial cycle.  Our results
therefore provide a complete taxonomy for generic Kerr orbits.

\bigskip
\bigskip
\bigskip 
\bigskip

**Acknowledgements**

This work was supported by an NSF grant AST-0908365. 
JL gratefully acknowledges
support of a KITP Scholarship, 
under Grant no. NSF PHY05-51164.

\bigskip
\bigskip

\vfill\eject

\appendix
\section{Spherical Orbits}
\label{sec:spherical}

In the cases of Schwarzschild and equatorial Kerr motion, orbits of
constant $r$ --- circular orbits --- serve to organize the ranges of
orbital parameters over which bound, nonplunging motion exists.
Constant $r$ orbits in the general Kerr geometry play a similar
organizational role but need not lie in a plane.  Thus, they are not
necessarily circular orbits but rather spherical.  Spherical orbits
were first treated in \cite{wilkins1972} and later analyzed in the
context of radiation reaction in \cite{hughes2001,hughes2001:2} (in
the latter references, these constant $r$ orbits are refered to as
``circular, nonequatorial orbits'', but we use the original shorter
moniker ``spherical'' from Ref.\ \cite{wilkins1972}).  Like circular
orbits, spherical orbits have $\dot{r}=\ddot{r}=0$; unlike their
circular counterparts, spherical orbits do not have $\dot{\theta}=0$.

An initial analysis of equatorial Kerr motion (we can think of
Schwarzschild motion as the $a=0$ subcase) begins with expressions for
$L_z$ and $E$ of circular orbits as a function of $r$ and the (fixed)
central black hole spin $a$.  Our generic Kerr analysis will reproduce
one such equatorial-like picture for each inclination $\iota$ and will
have an effective total angular momentum $L$ take the place of the
more conventional conserved quantity $L_{z}$ but otherwise proceed
analogously.  We therefore turn now to deriving expressions for the
effective angular momentum $L$ and $E$ of spherical orbits as
functions of $r, a$ and $\iota$.  Ref.\ \cite{hughes2001} has similar
expressions for $Q$ and $L_{z}$ of spherical orbits in terms of $r, a$
and $E$, but as we explain in Appendix \ref{sec:cons_quant},
aggregating orbits with fixed values of the constants $\iota$ and $L$
is most conducive to a clear exposition of the dynamics.

As in the equatorial Kerr case, our starting point is the radial
quasi-potential $R(r)$.  We begin by expressing $R(r)$ and its
derivatives in terms of $E, \iota$ and $L$.  From
eqn.\ (\ref{eq:Rpoly_iL}),
\begin{align}
  \label{eq:R}
  \begin{split}
  R\left(r\right) &= \left(E^2 - 1 \right) r^4 + 2r^3 + \left(a^2
  \left\{E^2 - 1\right\} - L^2 \right)r^2 \\
  &+ 2r \left(a^2 E^2 - 2 a E L\cos{\iota}
  + L^2 \right) + a^{2}L^{2}\left(\cos^{2}{\iota} -1 \right)   
  \end{split}
  \\
  \label{eq:R_prime}
  \begin{split}
  R'\left(r\right) &= 4\left(E^2 - 1 \right) r^3 + 6r^2+ 2\left(a^2 \left\{E^2 - 1\right\} - L^2 \right)r  \\
  &+ 2\left(a^2 E^2 - 2 a E L\cos{\iota}
  + L^2 \right)     
  \end{split}
  \\
  \label{eq:R_double_prime}
  R''\left(r\right) &= 12\left(E^2 - 1 \right) r^2 + 12r+ 2\left(a^2
  \left\{E^2 - 1\right\} - L^2 \right)
\quad .
\end{align}
The condition $\dot{r}=0$ implies $R\left(r\right)=0$ from equation
(\ref{subeq:dimcarter-r}).  Solving for $\ddot{r}$ from equation
(\ref{subeq:dimcarter-r}) we find that
\begin{eqnarray}
\label{eqn:ddotr}
\ddot{r} &=&
\frac{1}{2}\frac{\dot{R}}{\sqrt{R}}
\\
\nonumber
&=&
\frac{1}{2}\frac{\dot{r}R'}{\sqrt{R}}
\\
\nonumber
&= & \frac{1}{2} R'
\quad ,
\end{eqnarray}
where $R'\left(r\right)=\frac{dR}{dr}$.  We can see immediately from
equation (\ref{eqn:ddotr}) that $\ddot{r}=0$ implies
$R'\left(r\right)=0$.  Similarly, $\dddot{r}=0$ implies that
$R''\left(r\right)=0$.

To find expressions for all $E_{s}$ and $L_{s}$ for a fixed $a$ and
${\iota}$, we set $R\left(r\right)=R'\left(r\right)=0$ and solve for
$E_{s}\left(r,a,\iota\right)$ and $L_{s}\left(r,a,\iota\right)$.
Solving the two coupled quadratic equations yields four solutions for
each of $E_{s}$ and $L_{s}$.  We determine the physically admissible
solutions by imposing that $L_{s}$ always be positive, i.e.\ an
effective angular momentum \emph{magnitude}.  Additionally, because
each fixed ${\iota}$ should replicate the orbital structure of the
Schwarzschild geometry, both the $L_{s}$ and $E_{s}$
solutions should asymptote at low $r$-values to the innermost
time-like spherical orbit.  There should also be a minimum $L_{s}$ and
$E_{s}$ value corresponding to the innermost bound spherical orbit
(ibso).  And the $r$ at which the minima occur on the $L_{s}$ and
$E_{s}$ graphs should be the same.  Finally, at large $r$, our $L_{s}$
plot should reproduce the Newtonian limit, $\sqrt{L}\propto r$ and
$E_{s}$ should asymptote to $1$.

Combining the above conditions, we find
\begin{widetext}
\begin{align}
\begin{split}
E_{s}\left(r,a,\iota\right)
&=\biggl[\left(-3+r\right)\left(-2+r\right)^{2}r^{7}+a^{8}\sin^{4}\iota\left(1+r\right)
\\
& -2ar\cos{\iota}\Delta\left(-a^{2}\sin^{2}\iota+r^{2}\right)\sqrt{r\left(-a^{4}\sin^{2}\iota+2a^{2}\sin^{2}\iota\Delta+r^{4}\right)}
\\
& - a^{4}r^{2}\sin^{2}\iota\big[ a^{2}\left\{4-4\left(-1+r
\right)r+\cos^{2}\iota\left(1+r \right)\left(-5+4r \right)\right\} 
\\
& + 2\left(-1+r \right)r\left\{2-3\left(-2+r\right)r
+\cos^{2}\iota\left(-4+r\left(-1+2r \right) \right)\right\}\big]
\\
& + a^{2}r^{5}\big[4\left(-2+r \right)\left\{1+\left(-3+r \right)r
\right\} +\cos^{2}\iota \big\{8+r\left(-23+\left(17-4r \right)r
\right)\big\}\big]\biggr]^{\frac{1}{2}} 
\\
& /
\biggl[\left(-a^{4}\sin^{2}\iota-2a^{2}r^{2}\sin^{2}\iota-r^{4}
  \right) \times
\\
&\quad
  \left\{-\left(-3+r\right)^{2}r^{4}-a^{4}\sin^{2}\iota\left(1+r
  \right)^{2}+2a^{2}r^{2}\left(-\left(-3+r \right)\left(1+r
  \right)+\cos^{2}\iota\left(-3+r^{2} \right) \right) \right\} \biggr]^{\frac{1}{2}}
  \end{split}
\label{eq:E_spherical_kerr}
\\
L_{s}\left(r, a,\iota \right) &=
\frac{-\Delta\sqrt{r\left(-a^{4}\sin^{2}\iota+2a^{2}\sin^{2}\iota\Delta +
    r^{4}\right)}+ar\cos{\iota}\left(a^{2}+r\left(-4+3r\right)\right)}{-a^{4}\sin^{2}{\iota}-\left(-2+r\right)^{2}r^{2}+a^{2}r
\left(4-2r+\cos^{2}{\iota}\left(-3+2r\right)\right)} E_{s}\left(r,a,\iota\right)
\label{eq:L_spherical_kerr}
\end{align}
\quad .
\end{widetext}
We recover the functions $E_{c}$ and $L_{c}$ given in
\cite{bardeen1972} for equatorial Kerr circular orbits by setting
$\iota=0$ for prograde and $\iota=\pi$ for retrograde in equations
(\ref{eq:E_spherical_kerr}) and (\ref{eq:L_spherical_kerr}). From
there, we recover the well-known Schwarzschild functions $E_{c}$ and
$L_{c}$ (see, for instance, Ref.\ \cite{carroll}) by setting $a=0$ in
(\ref{eq:E_spherical_kerr}) and (\ref{eq:L_spherical_kerr}) (note
that, by spherical symmetry, those values must be and are independent
of $\iota$).

 \begin{figure}
  \includegraphics[scale=0.7]{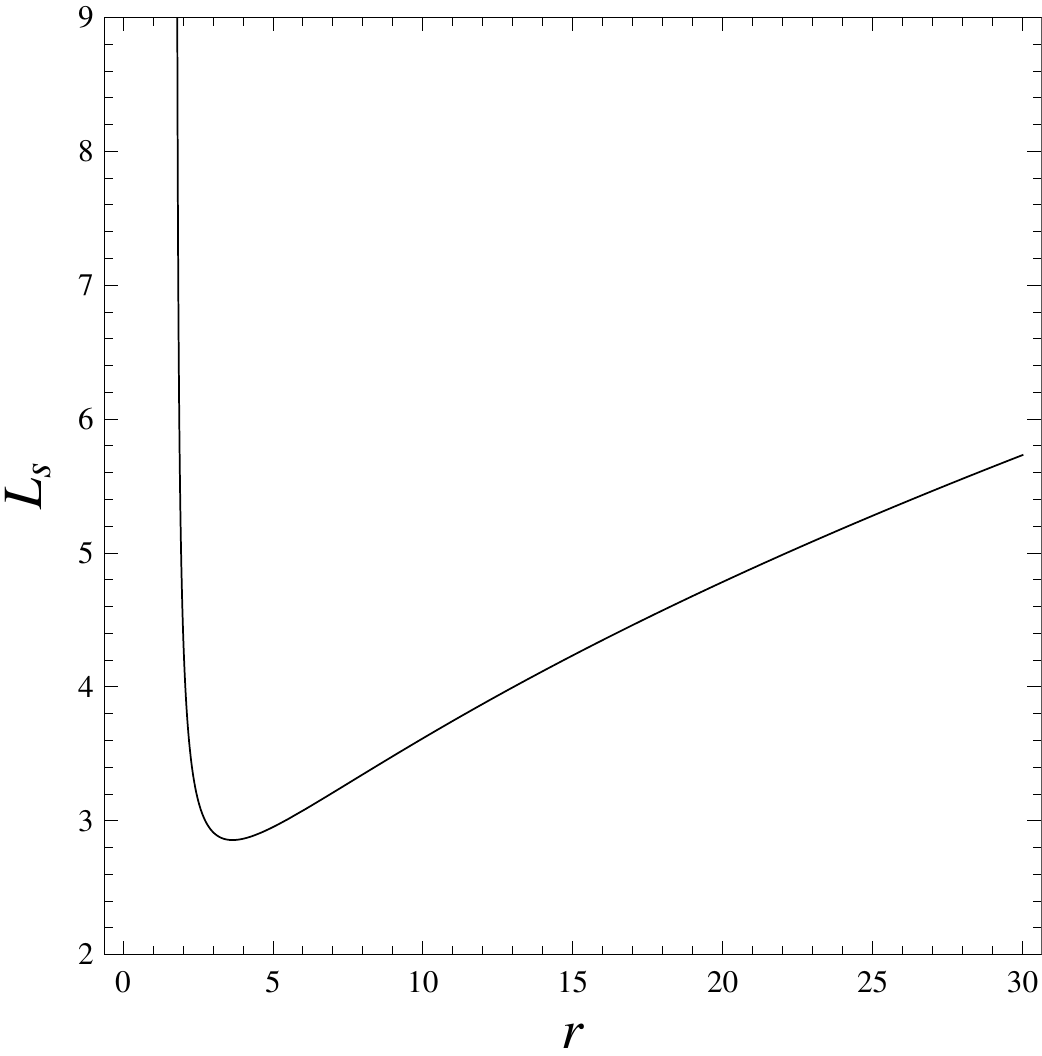}
  \includegraphics[scale=0.72]{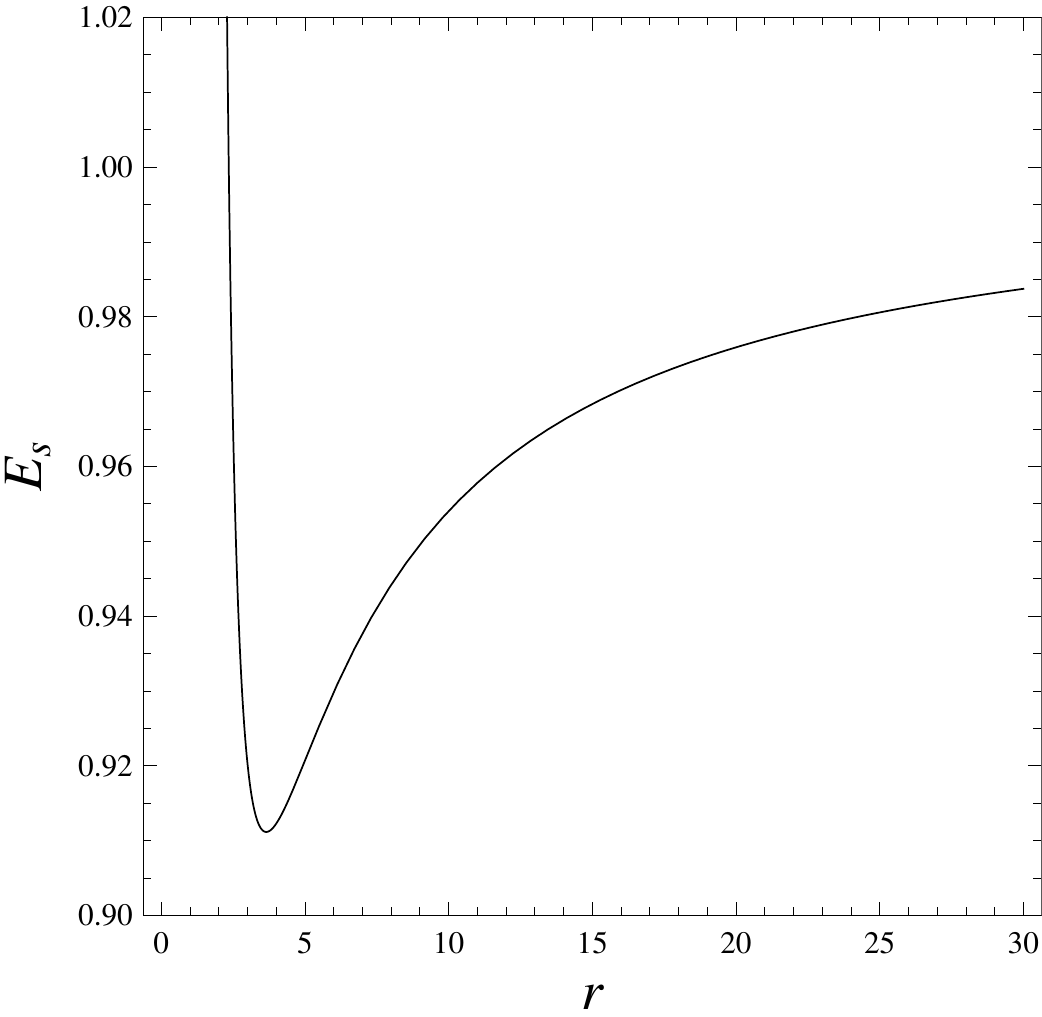}
  \caption{Top: The figure shows a plot of $L_{s}$ vs $r$ for
    spherical Kerr orbits with $a=0.99$ and $\cos{\iota}=0.4$.  Bottom:
    Shows a plot of $E_{s}$ vs $r$ for spherical Kerr orbits with the
    $a=0.99$ and $\cos{\iota}=0.4$.}
  \label{fig:L_E_r_kerr_noneq}
\end{figure}

Figure \ref{fig:L_E_r_kerr_noneq} shows both $L_{s}$ and $E_{s}$ as
functions of $r$ with parameters $\cos{\iota}=0.4$ and $a=0.99$.  The
following qualitative features are representative of all $\iota$ and
$a$ values and mimic the features of Schwarzschild.  Both $E_{s}$ and
$L_{s}$ have minima that occur at the same $r$.  The minimum $L_{s}$,
$L_{\text{isso}}$, corresponds to the least $L_{s}$ for which there
exists a spherical orbit.  The $V_{\rm eff}$ plot corresponding to
$L=L_{\text{isso}}$ has a saddle point where the stable and unstable
spherical orbits merge.  For all $L>L_{\text{isso}}$ there are two
spherical orbits, whose $r$-values exactly correspond to the local
minimum and maximum of the effective potential plots of that $L, \iota
\text{ and } a$.  The maximum is the unstable spherical orbit and the
minimum is the stable spherical orbit.  There is a critical value
$L_{s} =L_{\text{ibso}}$ at which the unstable spherical orbit has
$E_{s}=1$, and for all $L_{s}>L_{\text{ibso}}$, the unstable spherical
orbit is unbound with $E_{s}>1$.  For a fixed $\iota$ and $a$, all the
qualitative properties of the generic Kerr orbits replicate the
Schwarzschild system.

The innermost bound spherical orbit, $\text{ibso}$, is defined as the
spherical orbit with critical energy $E_{\text{ibso}}=1$.  To find the
$L_{\text{ibso}}$ and $r_{\text{ibso}}$, we set (\ref{eq:R}) and
(\ref{eq:R_prime}) to zero with $E=1$.  The innermost stable spherical
orbit, $\text{isso}$, is the minimum of the $L_{s}$ plot and is
subject to the further constraint $R''\left(r\right)=0$. We therefore
find the $\text{isso}$ for a given $\iota$ and $a$ by setting all
three of equations (\ref{eq:R}), (\ref{eq:R_prime}) and
(\ref{eq:R_double_prime}) to zero simultaneously and solving for
$L_{\text{isso}}$, $r_{\text{isso}}$ and $E_{\text{isso}}$.

\section{Choosing conserved quantities}
\label{sec:cons_quant}

The Kerr metric has four conserved quantities. They are conventionally
chosen to be the black hole mass ($\mu$), the orbital energy ($E$),
the $z$-component of angular momentum ($L_{z}$) and the carter constant
($Q$).  Because each of those quantities are constants of the motion,
any combination of them is also a constant of the motion.  Therefore,
there are an infinite number of choices of four independent quantities
we could make for our conserved quantities.

We have chosen to use $\mu$, $E$, effective angular momentum ($L$, where
$L=\sqrt{Q+L_{z}^2}$) and inclination angle ($\iota$, where
$\cos{\iota}=\frac{L_{z}}{L}$).  This section provides an explanation
for our choice.

Our goal was to realize a generic Kerr orbit structure that
generalized the Schwarzschild and equatorial Kerr orbit structures
presented in \cite{levin2008}.  To bring that goal to fruition, we
look for a set of conserved quantities such that we could hold one
fixed and reproduce all the qualitative features of Schwarzschild dynamics
($\text{isso}$, $\text{ibso}$, etc.).  

\begin{figure*}
  \centering 
\includegraphics[scale=0.74]{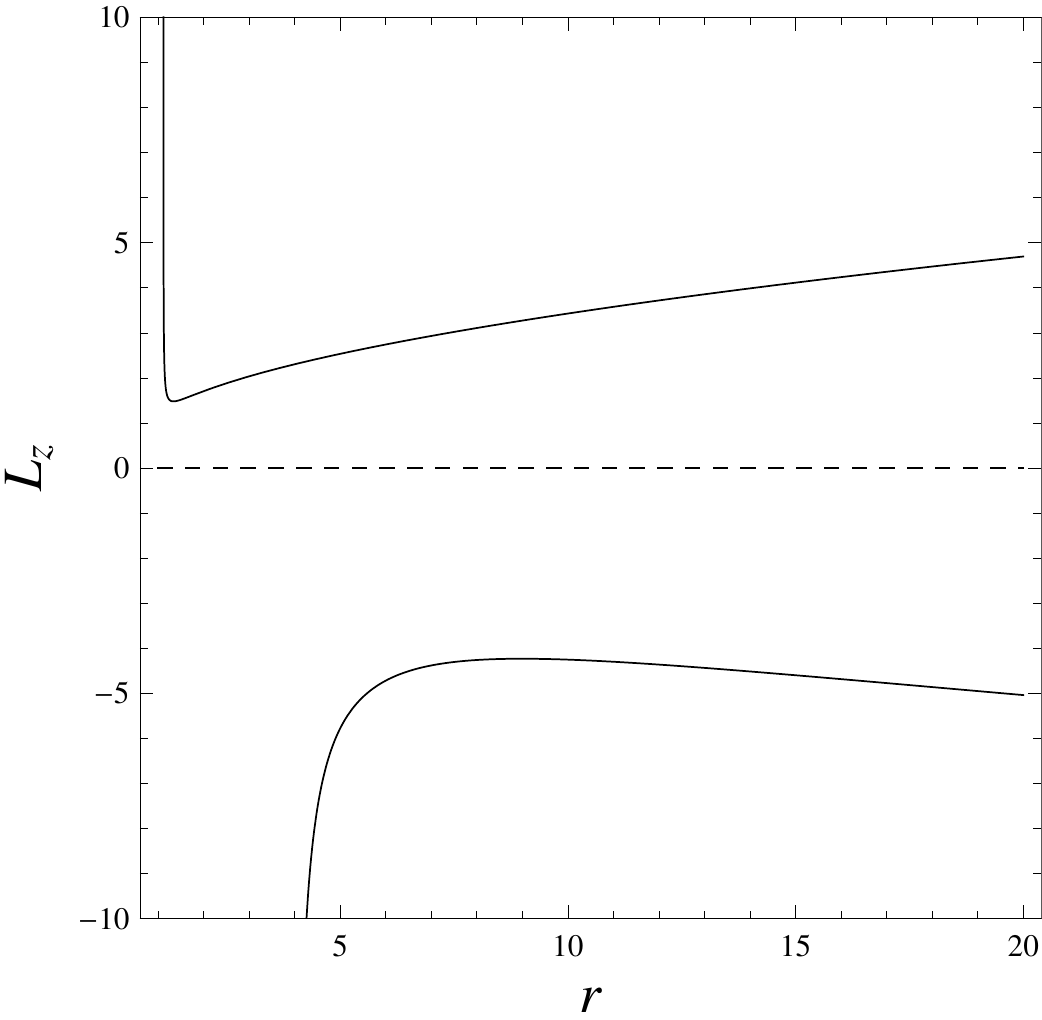}
  \includegraphics[scale=0.73]{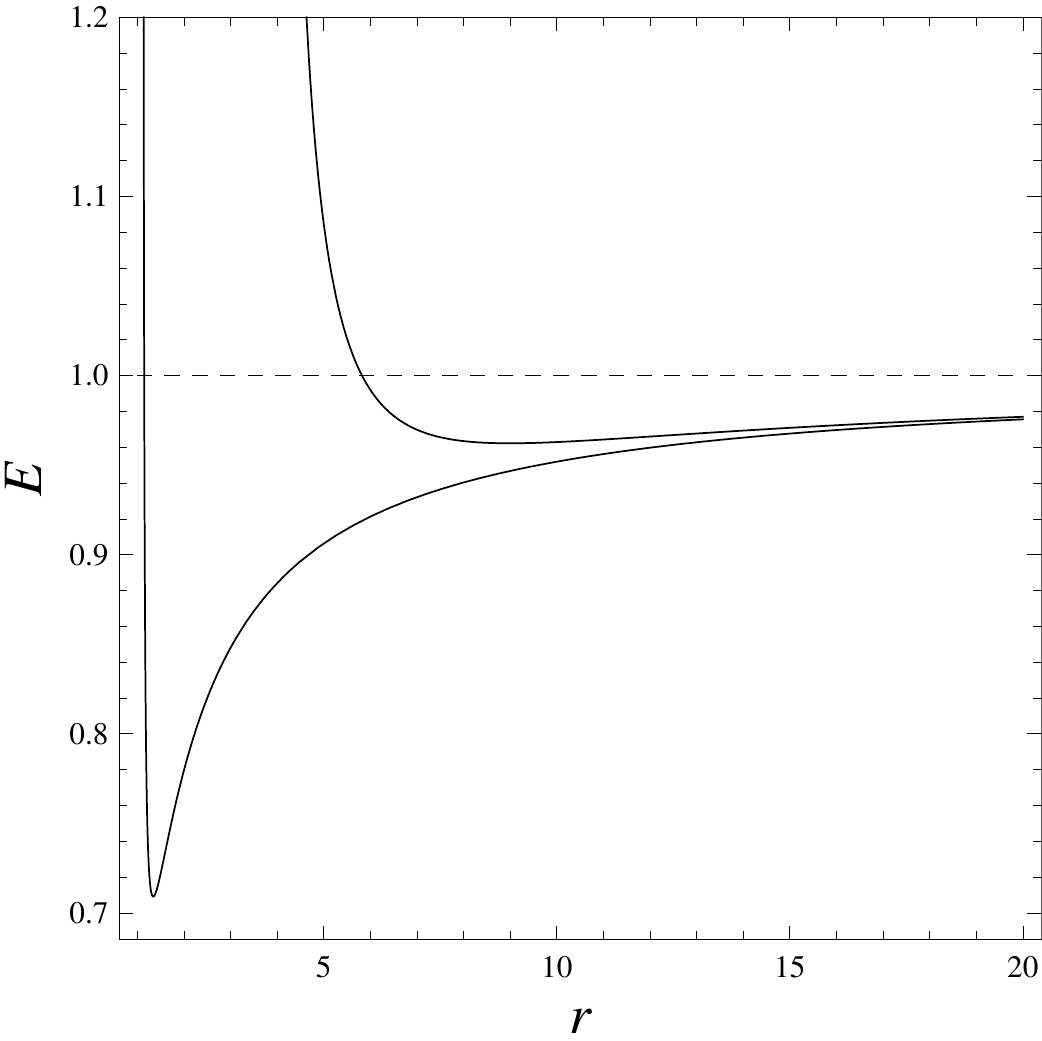}
  \caption{Left: The figure shows a plot of $L_{z}$ vs $r$ for
    equatorial circular Kerr orbits with $a=0.995$ and $Q=0$.  Right:
    Shows a plot of $E$ vs $r$ for circular Kerr equatorial orbits
    with $a=0.995$ and $Q=0$.}
  \label{fig:Q0LzE}
\end{figure*}

Using the conventional $Q$, $L_{z}$ and $E$, the equatorial Kerr
system is defined by $Q=0$.  There are two sets of $L_{z}$ and $E$
solutions for circular orbits, one prograde and one retrograde.
Figure \ref{fig:Q0LzE} shows the two solutions for $a=0.995$ and
$Q=0$.  We can see that the solutions never intersect and each
solution has all the qualitative features present in the standard
organization of Schwarzschild orbits.

However, when $Q$ becomes large enough, regardless of the spin, we see
a loss of adherence to these features.  Specifically, there is no
longer an $\text{isso}$, and the two sets of solutions for $L_{z}$ and
$E$ for a fixed $Q$ mix.  While this phenomenon is not seen until $Q$
gets large, it is present for all spin values.  The discontinuity in
the $L_{z}$ and $E$ spherical graphs, as well as the loss of the
$\text{isso}$ is seen for the full range of $a$ values.

The upshot is that there are values of $Q$ that do not allow us to
reproduce the familiar qualitative organization of Schwarzschild
dynamics if we choose to look at orbits of constant $Q$ as an
ensemble.  In contrast, we find that with $(E,L,\iota)$, for every
fixed $\iota$, the qualitative dynamical picture mimics the familiar
Schwarzschild one beautifully.  In this picture, each $\iota$
corresponds to a fixed orbital inclination so that equatorial orbits
correspond to one of two $\iota$ values: $\iota=0$ for prograde, and
$\iota=\pi$ for retrograde.  Furthermore, whereas each fixed $Q$
admits two associated $E$ and $L_{z}$ solutions each for spherical
orbits, each $\iota$ produces only one curve each for $L_{s}$ and
$E_{s}$.

\begin{figure*}[hb]
\includegraphics[scale=0.67]{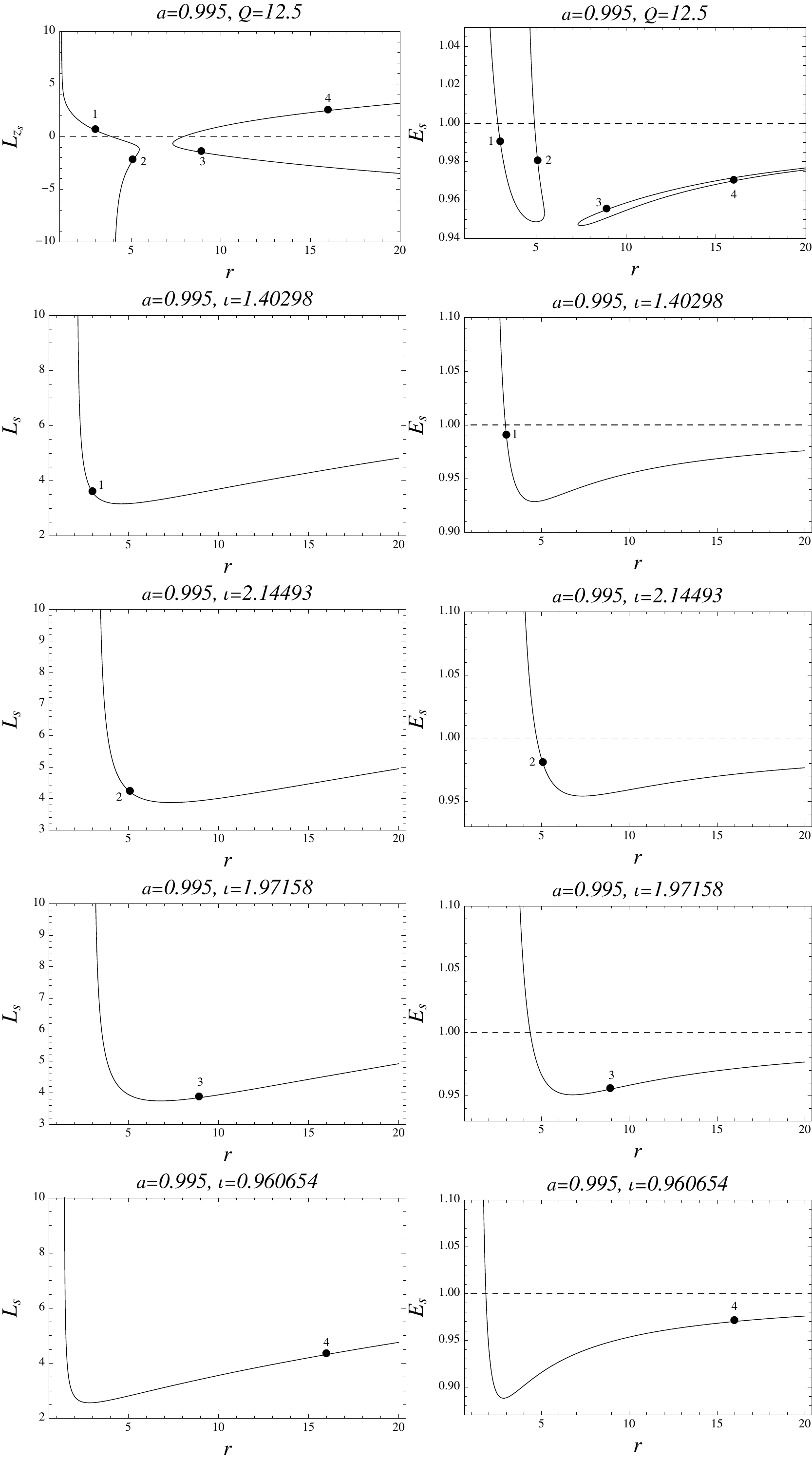}
\caption{The above pictures show the organizational differences
  between using the conserved quantities $E$, $L_{z}$ and $Q$ and
  using $E$, $L$ and $\iota$.  All plots are for spherical orbits with
  $a=0.995$.  Top: Curves of $L_{z_{s}}$ vs.\ $r$ and $E_{s}$ vs.\ $r$
  for spherical orbits all with fixed $Q=12.5$.  Below: Curves of
  $L_{s}(r)$ and $E_{s}(r)$ for spherical orbits with the four
  different fixed $\iota$ values associated with the four labeled
  points. Each such point corresponds to the same orbit in both the
  fixed $Q$ and fixed $\iota$ graphs.  Their parameter values are: (1)
  $E=0.99$, $L_{z}=0.598971$, $r=3.01492$, $L=3.58591$,
  $\iota=1.40298$; (2) $E=0.98$, $L_{z}=-2.28682$, $r=5.09346$, $L=
  4.21065$, $\iota=2.14493$; (3) $E=0.955$, $L_{z}=-1.49806$, $r=
  8.92632$, $L=3.83981$, $\iota=1.97158$; (4) $E=0.97$,
  $L_{z}=2.47180$, $r=15.9948$, $L=4.31391$, $\iota=0.960654$.
  Orbits 1 \& 2 are unstable; 3 \& 4 are stable.}
\label{fig:ELpts}
\end{figure*}

Figure \ref{fig:ELpts} shows a set of $E$ and $L_{z}$ plots for
spherical orbits with $Q=12.5$.  We can see the loss of the
$\text{isso}$ and the mixing of the two seperate solutions.  The
curves are no longer even single-valued at a given $r$.  Moreover, the
$E_{s}(r)$ curve can have more than 2 orbits with a given $E$, as
opposed to only the stable and unstable constant $r$ orbits we are
used to in the Schwarzschild effective potential picture.  We have
picked four points on the fixed $Q$ plots, each with a unique set of
orbital parameters, $E$, $L_{z}$ and $Q$.  For each of those points,
we have determined the corresponding $E$, $L$ and $\iota$ and plotted
the $E_{s}(r)$ and $L_{s}(r)$ curves for each of those $\iota$ values.
Notice that there is no such breakdown when we look at curves of fixed
$\iota$ rather than fixed $Q$.  Instead, the latter curves faithfully
reproduce the expected qualitative features of the corresponding
Schwarzschild or equatorial Kerr curves.

\bibliography{kerr_tax}

\begin{thebibliography}{10}

\bibitem{carter1968}
B.~Carter,
\newblock Phys. Rev. {\bf 174}, 1559 (1968).

\bibitem{misner}
C.~W. Misner, K.~S. Thorne, and J.~A. Wheeler,
\newblock {\em Gravitation},
\newblock W. H. Freeman, first edition, 1973.

\bibitem{levin2008}
J.~Levin and G.~Perez-Giz,
\newblock Phys. Rev. D {\bf 77}, 103005 (2008).

\bibitem{levin2009}
J.~Levin,
\newblock Class. Quant. Grav. {\bf 26}, 235010 (2009).

\bibitem{levin2008:2}
J.~Levin and R.~Grossman,
\newblock gr-qc/08093838  (2008).

\bibitem{grossman2008}
R.~Grossman and J.~Levin,
\newblock gr-qc/08113798 .

\bibitem{mino2003}
Y.~{Mino},
\newblock \prd {\bf 67}, 084027 (2003).

\bibitem{wilkins1972}
D.~C. Wilkins,
\newblock Phys. Rev. {\bf D5}, 814 (1972).

\bibitem{levin2008:3}
{J. Levin and G. Perez-Giz},
\newblock {\em {Homoclinic Orbits around Spinning Black Holes I: Exact Solution
  for the Kerr Separatrix}},
\newblock 2008.

\bibitem{hughes2001}
S.~A. Hughes,
\newblock erratum-ibbid.d {\bf 63}, 049902 (2001).

\bibitem{drasco2006}
S.~Drasco and S.~Hughes,
\newblock Phys. Rev. D {\bf 73}, 024027 (2006).

\bibitem{schmidt2002}
W.~Schmidt,
\newblock Class.\ Quant.\ Grav. {\bf 19}, 2743 (2002).

\bibitem{drasco2004}
S.~{Drasco} and S.~A. {Hughes},
\newblock Phys.\ Rev.\ D {\bf 69}, 044015 (2004).

\bibitem{ryan1995}
F.~D. Ryan,
\newblock Phys. Rev. {\bf D52}, 3159 (1995).

\bibitem{ryan1996}
F.~D. Ryan,
\newblock Phys. Rev. {\bf D53}, 3064 (1996).

\bibitem{hughes2001:2}
S.~A. Hughes,
\newblock Phys.\ Rev.\ D {\bf 64}, 064004 (2001).

\bibitem{glampedakis2002:2}
K.~Glampedakis, S.~A. Hughes, and D.~Kennefick,
\newblock Phys.\ Rev.\ D {\bf 66}, 064005 (2002).

\bibitem{perez-giz2008}
{G. Perez-Giz and J. Levin},
\newblock {\em {Homoclinic Orbits around Spinning Black Holes II: The Phase
  Space Portrait}},
\newblock {2008}.

\bibitem{carroll}
S.~Carroll,
\newblock {\em Spacetime and Geometry: An Introduction to General Relativity},
\newblock Benjamin Cummings, 2003.

\bibitem{bardeen1972}
J.~M. {Bardeen}, W.~H. {Press}, and S.~A. {Teukolsky},
\newblock Ap.\ J. {\bf 178}, 347 (1972).

\end{thebibliography}
\bibliographystyle{aip}

\end{document}